\documentclass[useAMS,usenatbib]{mn2e}

\usepackage{graphicx}
\usepackage{natbib}
\usepackage{lscape}

\title[Variables in M54]
{The population of variable stars in M54 (NGC6715)\thanks{Based on 
observations taken at the ESO-Danish 1.54-m telescope in La Silla, Chile}}
\author[Sollima et al.]{A. Sollima$^{1}$, 
C. Cacciari$^{2}$, M. Bellazzini$^{2}$ and S. Colucci$^{3}$\\   
$^{1}$Instituto de Astrofisica de Canarias, C/Via Lactea s/n, E-38205, 
San Cristobal de La Laguna, Tenerife, Spain; \\ 
e-mail:asollima@iac.es \\
$^{2}$INAF Osservatorio Astronomico di Bologna, via Ranzani 1,
Bologna, 40127-I, Italy    \\
$^{3}$ Formerly at the Dipartimento di Astronomia, Universit\`a di Bologna, Via Ranzani 1, 
I-40127, Bologna, Italy}
\begin{document}

\date{Accepted 2010, ???; Received 2010, ???; in original form
2010, ???}

\pagerange{\pageref{firstpage}--\pageref{lastpage}} \pubyear{2009}

\maketitle

\label{firstpage}

\begin{abstract}
We present new B, V and I CCD time-series photometry for 177 variable stars 
in a $13\arcmin\times 13\arcmin$ field centered on the globular cluster M54 
(lying at the center of the Sagittarius dwarf spheroidal galaxy), 94 of which are 
newly identified variables. 
The total sample is composed of 2 anomalous Cepheids, 
144 RR Lyrae stars (108 RR0 and 36 RR1), 3 SX Phoenicis, 
7 eclipsing binaries (5 W UMA and 2 Algol binaries), 3 variables of uncertain 
classification and 18 long-period variables. The large majority of the RR Lyrae 
variables likely belong to M54. Ephemerides are provided for all the 
observed short-period variables.
The pulsational properties of the M54 RR Lyrae variables are  
close to those of Oosterhoff I clusters, but a significant 
number of long-period ab type RR Lyrae are present. 
We use the observed properties of the RR Lyrae to estimate the reddening and the 
distance modulus of M54,
E(B-V)=0.16 $\pm$ 0.02 and (m-M)$_0$=17.13 $\pm$ 0.11, respectively, in excellent 
agreement with the most recent estimates. 
The metallicity  has been estimated for a subset of 47 RR Lyrae stars, with 
especially good quality light curves,
from the Fourier parameters of the V light curve. The derived metallicity 
distribution has a symmetric bell shape, with a mean of 
$\langle [Fe/H]\rangle= -1.65$ and a standard deviation $\sigma= 0.16$ dex. 
Seven stars have been identified as likely belonging to the Sagittarius galaxy, 
based on their too high or too low metallicity. 
This evidence, if confirmed, might suggest that old stars in this galaxy span a 
wide range of metallicities. 

\end{abstract}


\section{Introduction}\label{s:int}

The globular cluster M54 (NGC6715) is a special object under several aspects.
In fact, it is quite massive ($\sim 2\times 10^6~M_{\sun}$, Pryor \& Meylan 
1993), it displays a complex Horizontal Branch (HB) morphology (extending up to 
the highest temperatures achievable by these stars; Rosenberg et al. 2004) and 
its stars present an intrinsic spread in Iron abundances (Carretta et al. 2010, 
C10 hereafter; see Bellazzini et al. 2008, hereafter B08, for previous studies 
and references). 
Moreover, it resides at the center of the Sagittarius dwarf spheroidal galaxy 
(Sgr dSph; Ibata, Gilmore \& Irwin 1994), a satellite of the Milky Way that is 
currently disrupting under the tidal strain of the Galaxy (see, for example, 
Majewski et al. 2003). Several authors suggested the possibility that M54 could 
be the nucleus of Sgr (Bassino \& Muzzio 1995; Sarajedini \& Layden 1995). 
Later  studies (Layden \& Sarajedini 2000; Majewski et al. 2003; 
Monaco et al. 2005a), however, demonstrated that the metal-poor cluster 
([Fe/H]$\simeq -1.55$; Brown, Wallerstein \& Gonzalez 1999, C10) co-exists with 
a stellar nucleus made of metal rich stars of the Sgr galaxy 
($\langle[Fe/H]\rangle\simeq -0.45$; B08 and references therein).
A recent study based on accurate velocity and metallicity 
data (B08) strongly supports the idea that M54 formed 
independently, and plunged into the Sgr nucleus as a result of 
significant decay of its original orbit due to dynamical friction (Monaco et al.
2005b; 
see also C10 for independent support to this hypothesis). 

In this context, RR Lyrae (RRL) variables can provide further insight into the 
nature of this stellar system, in addition to reliable estimates of the distance 
and foreground extinction. 
The first (photographic) survey for variable stars in M54 was 
performed by Rosino \& Nobili (1959) who identified 82 variables.
 Their pioneering analysis was followed 
up and expanded by Layden \& Sarajedini (2000; LS2000 hereafter), 
who considered the variables as part of their V,I CCD photometric 
study aimed at obtaining the colour-magnitude diagram (CMD) of M54 + Sgr field 
to derive the star formation history of the Sgr dSph galaxy.
In that work LS2000 found 117 variable stars, 93 of which are candidate 
RRL stars. They analysed the best observed subset of RRL (67 stars) to 
obtain a distance modulus of $(m-M)_0=17.19\pm0.12$ and E(V-I)=$0.18\pm 0.02$. 
On the basis of the pulsational properties of their sample of RRL, these authors 
classified M54 as an  Oosterhoff type I cluster (Oosterhoff 1939; see Catelan 
2009 for a recent review). 

In this paper we present new BVI photometric data and discuss in detail the 
characteristics of the RRL stars of M54, that were presented in preliminary 
form by Cacciari, Bellazzini \& Colucci (2002). 
In Sect. 2 we describe our data, the observations, the data 
reduction, the calibration procedure and the period search procedure. 
In Sect. 3 we present and briefly discuss the
characteristics of the CMD resulting from our photometric data and present our 
sample of detected variables. In Sect. 4 we analyze the 
RRL pulsational properties and use them to estimate their membership to the M54 
or Sgr stellar population. Sect. 5 presents the results of the Fourier analysis 
of the RRL light curves and metallicity estimates.  
The reddening and distance are derived in Sect. 6, and 
the properties of the other variables of our sample are briefly presented and 
discussed in Sect. 7. We summarize and discuss our results in Sect. 8. 
In Appendix A we comment on individual variables.

\section{The data}\label{s:data}

\subsection{Observations}\label{s:obs}

The BVI photometric data presented in this paper were obtained on 13, 14 
and 19 July 1999 at the ESO-Danish 1.54-m telescope in La Silla (Chile) 
using the 2k$\times$2k CCD of the Danish Faint Object Spectrograph and Camera 
(DFOSC). 
The images cover a $\simeq 13\arcmin\times13\arcmin$ field of view roughly centered on 
M54, with a scale of 0.39 $\arcsec$/pixel. The BVI images were taken 
sequentially with exposure times of 360-900 sec in B,  
150-600 sec in V, and 150-450 sec in I, depending on the atmospheric 
conditions. The average seeing during the three nights ranged between 
$1.9\arcsec$ and $2.5\arcsec$ (FWHM).
In total, it was possible to obtain 54 B, 57 V and 52 I images. 

\subsection{Data reduction}\label{s:dr}

The frames were overscan corrected, bias subtracted and flat fielded (using 
twilight sky flats) by means of the standard IRAF\footnote{IRAF 
is distributed by NOAO, which are operated by the Association of Universities 
for Research in Astronomy Inc., under cooperative agreement with the National 
Science Foundation (NSF)} tasks. 

The data reduction was performed using the ISIS package (Alard 2000), which 
is based on the method of image subtraction. ISIS is able to transform a series 
of images of the same field to (a) the same astrometric system, (b) the same 
flux scale, and (c) the same Point Spread Function (PSF)\footnote{In this task 
one of the original images is taken as {\em astrometric reference}. Usually the image 
with the best seeing and the lowest background level is selected for this 
purpose (see Alard 2000).}, without flux losses. Once all the above 
transformations have been performed, the data reduction proceeds as follows:

\begin{enumerate}
\item A high signal-to-noise reference frame (per filter) is constructed, by 
stacking a suitable number of images; in the present case the reference 
frames were built by stacking 14 B, 15 V and 14 I frames.
 
\item Each individual image is subtracted from the reference frame; the squares
of the images of the residuals are co-added into a single image where there are only 
the variable sources.

\item Variable stars are identified on the stacked images of the residuals and 
their positions are recorded.

\item The residual flux is measured on each single subtracted image at the 
position of the detected variables, using a PSF fitting routine. 
A light curve in units of flux difference is obtained for each variable.

\item Photometry of the reference frames is obtained independently (in our case 
using DoPHOT; Schechter, Mateo \& Saha 1993) to provide the flux zero point
to the light curves (and convert into a scale of magnitudes). In our case we 
transformed the instrumental magnitudes obtained with DoPHOT into absolute 
Johnson-Cousins magnitudes using several hundreds of stars in common with the 
photometry by LS2000 (for V and I) and by 
Rosenberg, Recio-Blanco \& Garcia-Marin (2004) for B magnitudes.
\end{enumerate}

\begin{figure}
 \includegraphics[width=8.7cm]{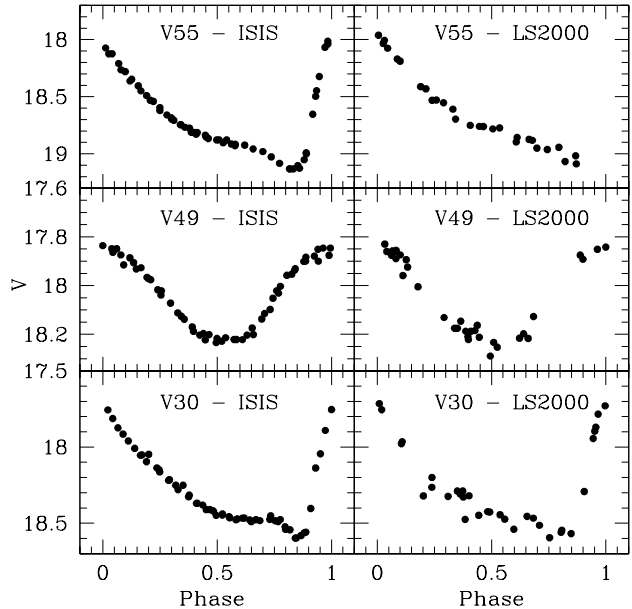}
\caption{Comparison between the light curves obtained in this wprk
(ISIS reduction package) and by LS2000 (DoPHOT) for three common stars of M54.}
\label{f:compar}
\end{figure}

The image-subtraction technique has proven very powerful in detecting even low 
amplitude variables in crowded fields\footnote{With this technique variables 
can be detected even in un-resolved portions of stellar fields images, as the 
unresolved bulk of constant stars is subtracted out, leaving only the isolated 
variables on the subtracted images.}, ensuring a higher accuracy in relative 
photometry with respect to traditional techniques 
(see e.g. Olech et al. 1999; 
Mochejska et al. 2004; Corwin et al. 2006; Greco et al. 2009; we refer to 
these papers also for further details on the technique).
As an example, in Fig. \ref{f:compar} we show for comparison the V light 
curves of 3 RRL variables in M54 that were measured by LS2000 with 
traditional techniques (DoPHOT) and by us with ISIS: the higher quality of the 
new light curves is readily evident.

There are two additional features of the specific application of the method performed in the 
present case that deserve to be described: 
\begin{itemize}
\item The images of a few bright foreground stars were saturated in our
 frames. In the process of image subtraction these sources left spurious 
 residuals due to different degree of saturation in the different frames, 
 blooming etc. This in turn led to the identification of large numbers of 
 spurious variable sources in the surroundings of these saturated stars. 
 To remove contaminants from the list of variables before obtaining their light 
 curves, we masked the surroundings of heavily saturated stars. 
 The dimension of the masked areas was decided case by case, by inspecting the 
 stacked image of the residuals.
\item DoPHOT automatically masks the regions surrounding (even moderately) 
saturated stars\footnote{The level of saturation at which masking is invoked 
depends also on the user choices. We adopted a conservative approach here, to 
keep only well measured stars in our DoPHOT catalog.} and/or sources for which 
the PSF fit does not reach convergency. For this reason there are variable 
stars for which we obtain a good differential flux light curve from ISIS but 
are not measured by DoPHOT on the reference image. 
Hence they lack a zero-point and their light curves cannot be transformed into 
a magnitude scale. For these stars we cannot measure an absolute amplitude, 
but we can obtain reliable periods and classify them on the basis of their
periods, the shape of the light curve and the similarity of the differential 
flux amplitudes with calibrated variables.
\end{itemize}

\subsection{Light curve analysis}\label{s:lc}

For the period search, we used the Graphical Analyzer of Time Series (GrATiS) 
program (see Clementini et al. 2000 for details), which was run on the ISIS 
differential flux data. This code (developed by P. Montegriffo at the 
Osservatorio Astronomico di Bologna) employes two different algorithms: (1) a 
Lomb periodogram (Lomb 1976) and (2) a best fit of the data with a truncated 
Fourier series (Barning 1963). The adopted period search procedure 
performs the Lomb analysis on a wide period interval. 
The Fourier algorithm refines the period definition and find the best-fitting 
model. 
The r.m.s. of the observed points about the best fit model is taken as
the error associated to the mean magnitudes. This value is reported along with 
the photometric data for each variable in 
Table 1\footnote{Table 1 is available in its entirety  
in the electronic edition}.

\subsection{Astrometry}

The x,y coordinates in pixels have been transformed into J2000 Equatorial 
coordinates with a second order polynomial fitted to $\sim 2000$ stars in 
common with 2MASS (Skrutskie et al 2006), that were used as astrometric 
standards. The residuals of the adopted astrometric solution were 
$\la 0.14\arcsec$ r.m.s. in both RA and Dec. 

\section{Photometric results} 

\subsection{Colour Magnitude Diagram}\label{s:cmdo}

\begin{figure}
 \includegraphics[width=8.7cm]{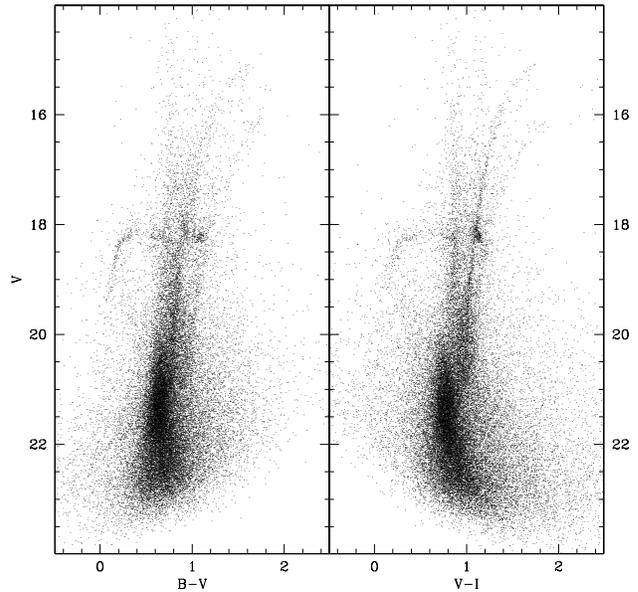}
\caption{V-(B-V) (left panel) and V-(V-I) (right panel) CMDs of M54 and Sgr, 
based on the present data. }
\label{cmd}
\end{figure}

In Fig. \ref{cmd} the V-(B-V) and V-(V-I) CMDs obtained with DoPHOT from the 
reference BVI images are shown.
The CMD reaches a limiting magnitude of $V\sim23$ sampling the 
population of M54 and Sgr down to $\sim$ 2 magnitudes below the cluster 
main sequence  turn-off.
The vertical plume of stars at $B-V(V-I)\ga 0.6(0.8)$ is composed by 
foreground/background Galactic main sequence stars at various distances along the 
line of sight. The steep and narrow Red Giant Branch (RGB) of M54 is clearly 
visible, bending from $B-V(V-I)\simeq 0.8(1.0)$ at $V\simeq 20$ to 
$B-V(V-I)\simeq 1.6(1.8)$ at $V\simeq 15$. The wider (and sparser) RGB of Sgr 
runs nearly parallel to the red of the M54 one. 
A well-defined blue HB (mostly associated to M54; see Monaco et al. 2004)  and 
a well populated Red Clump (associated to Sgr) are also visible at V$\sim$18.2.
A detailed discussion of the CMD is out of the scope of the present paper; we 
address the interested reader to papers more focused on the interpretation of 
the CMD (Sarajedini \& Layden 1995; Monaco et al., 2003, 2004, and 
references therein). For a detailed analysis of the stellar content in the 
innermost region of the cluster, see the exquisite CMD obtained by Siegel et 
al. (2007) from Hubble Space Telescope data.
For the purposes of the present analysis, it is important to 
recall that within the limiting radius of M54 ($r_t=6.3\arcmin$; see Ibata et 
al. 2009), the surface density of the cluster is nearly everywhere a factor of 
7-8 larger than the Sgr metal-rich nucleus (B08). Moreover, the population of 
the Sgr nucleus is dominated by intermediate-age stars that are unlikely to 
evolve into RRL (Siegel et al. 2007 and references therein) producing a
fraction of RRL which should be much lower than in M54 (Monaco et al. 2004).
This is in agreement with Cseresnjes (2001), who estimated a density of
139 RR0 per square degree in the central region of Sgr, leading to about 6.5 RR0 in our
13 arcmin square field of view.
Therefore, we conclude that the sample of RR Lyrae assembled in the present 
study must be dominated by members of M54. 

\subsection{The Variables}\label{s:var} 

\begin{figure}
\includegraphics[width=8.7cm]{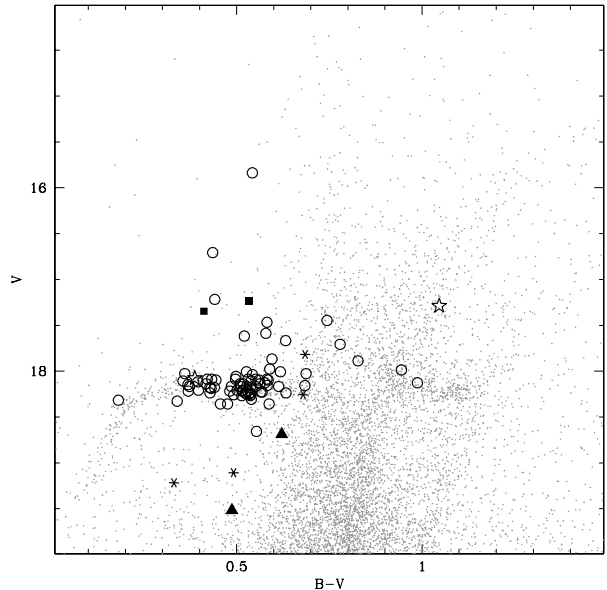}
\caption{Expanded V-(B-V) CMD of M54. The detected variables are shown as: 
open circles  (RR Lyrae stars), filled squares  (Cepheids), asterisks 
(eclipsing binaries), filled triangles  (SX Phoenicis) and open stars (uncertain
classification).}
\label{cmdvar}
\end{figure}

Our photometric search allowed us to identify 177 variable stars 
(94 new discoveries), significantly increasing the number (117) of variables 
identified by the previous studies of Rosino \& Nobili (1959) and LS2000. 
The identification and photometric information are given in Table 1.
In Fig. \ref{cmdvar} the variables detected in this study are identified in  
the V-(B-V) CMD.

Among the 89 stars studied by LS2000: 2 (V23 and V91) were not identified in
the present study because they lie out of our FoV; 4 (V25, V108, V111 and V112)
were not identified because they were masked out during the variable detection phase (see
Sect. \ref{s:dr}); 4 (V86, V89, V90 and V95) were identified as variables but 
no light curve could be derived and one (V100) was found to be non
variable. 
Of the variables reported in the former study of Rosino \& Nobili (1959) 
5 of them (V21, V26, V73, V79 and
V81) were not identified, neither by us nor by LS2000, because no stars 
were detected at the locations indicated by Rosino \& Nobili (1959) and 5 
(V20, V22, V27, V53 and V72) were found to be non variable (as also 
found by LS2000). These stars are not listed in Table 1. 
Finally, 18 stars were identified and analysed using only internally calibrated 
photometry; they do appear in Table 1 but no photometric information
is provided. The agreement between the light curves and the estimated periods 
between us and LS2000 is very good, in general. A few discordant cases are discussed in Appendix A.

For the 61 variable stars of the LS2000 sample for which we have 
calibrated photometry, we obtained light curves and derived 
the ephemerides using LS2000 photometric data in combination with ours to 
obtain a longer time baseline and more accurate results. 

For the 94 newly identified variables (V118-V211) we derived the classification 
and the periods. 
We found: 51 RR0, 29 RR1, 3 SX Phoenicis, 3 W Uma (EW), one Algol
binary (EA), 4 long
period (LP) candidates (with no period determination) and 3 with uncertain 
classification (V145, V147 and V211) of which two (V147 and V211) might be 
eclipsing binaries (EB). From the present calibrated data we find no evidence
of double-mode pulsation in any of the studied stars.
For 58 of these stars, mostly located in the central region of M54, 
the photometry could not be externally calibrated for the reasons described in 
Sect. \ref{s:dr}. 

\begin{figure*}
 \includegraphics[width=12cm]{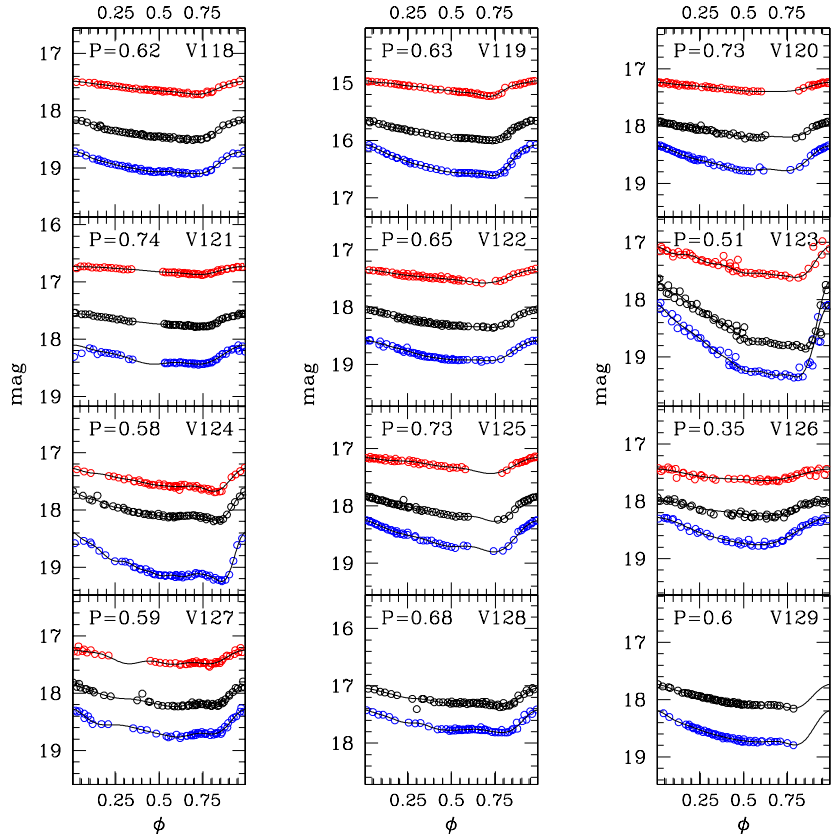}
\caption{Light curves of a sample of newly discovered RR Lyrae variables of M54. 
Blue (lower), black (middle) and red (upper) data points indicate B,V and I magnitudes, respectively. 
The best fit models obtained with GrATIS are overplotted.}
\label{lc_rrlyr}
\end{figure*}

\begin{figure*}
 \includegraphics[width=12cm]{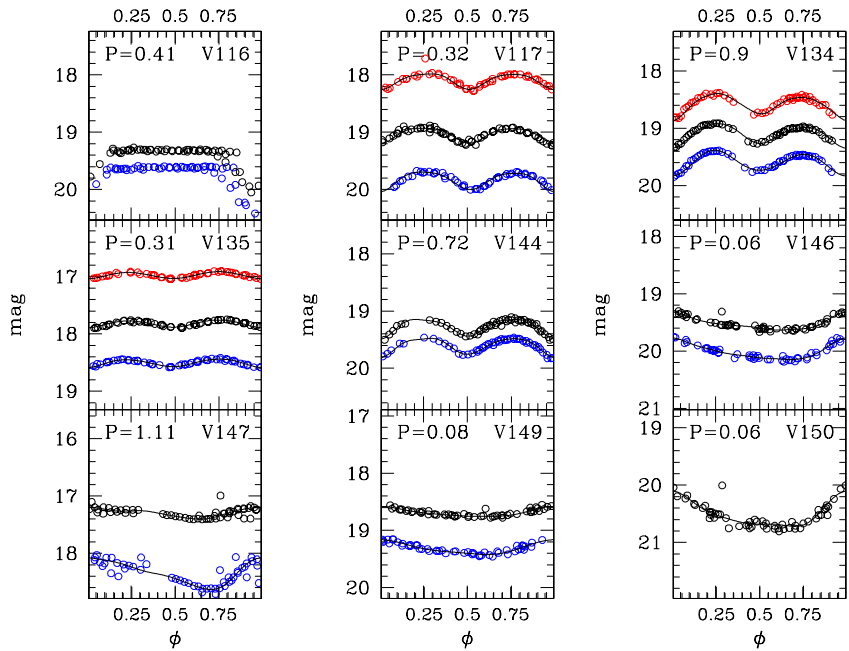}
\caption{Same as Fig. \ref{lc_rrlyr} for other types of variables of M54.}
\label{lc_other}
\end{figure*}

The light curves of a sample of newly discovered RRL and other variables 
are shown in Fig.s \ref{lc_rrlyr} and \ref{lc_other}.\footnote{The 
complete sample of light curves is available in the electronic edition.}
The epoch photometry for all the variables with calibrated photometry is available in its 
entirety at the CDS\footnote{http://vizier.u-strasbg.fr/cgi-bin/VizieR}. 

In the following we will focus the analysis only on RR Lyrae stars; variables of 
other types will be briefly discussed in Sect.~\ref{s:other}.

\begin{landscape}
\begin{table}
\label{t:all}
 \centering
 \begin{minipage}{225mm}
  \caption{The RR Lyrae variable stars in the M54 field of view: a sample of the online catalog 
  (the full catalog is available in the electronic version).  
  Columns are: (1) Identification (ID) from Clement et al. (2001) and LS2000 up to V117, and from this paper for 
  V118 to V221; (2) variable types: 0=Cepheid; 1=RR0; 2=RR1; 3=SX Phe;
  4=LP; 5=EW; 6=EA; 7=Uncertain (3-4) RA and Dec coordinates. The epoch of the astrometric reference is J2000; (5) Pulsation period (6) Epoch of maximum light; (7-9) mean
  BVI magnitudes calculated as intensity averages over the pulsation cycle converted back to magnitudes; (10)  r.m.s. of the observed V light curve points about the best fit 
  model; (11-13) BVI amplitudes; (14-15) B-V and V-I colors at minimum light (phase interval 0.5-0.8); (16-17) mean $<B-V>$ and $<V-I>$ colors calculated as averages over 
  one pulsation cycle of the best fit models; (18) Notes: Bl = possible Blazhko variability; b = possible blend; Cal =
  calibration problem; F = possible
  Galactic field member; Sgr = possible Sgr member; Ev = possible evolved.}
  \begin{tabular}{@{}lccccccccccccccccl@{}}
  \hline
   ID & Ty	& RA (J2000)  & Dec (J2000)  & Period	   & Epoch & $B$   & $V$   & $I$   & rmsV & $A_B$& $A_V$& $A_I$ & (B-V$)_{min}$ & (V-I$)_{min}$ &
   $<$B-V$>$ & $<$V-I$>$ & Note \\
  n.  & 	& deg	      & deg	     & d	   & HJD      &       &       &       &      &      &	   &	  &	&    &  &  &  \\
      & 	&	      & 	     &  	   & 2451370+ &       &       &       &      &      &	   &	  &	&    &  &  &  \\
 \hline \\
   01 &  0	& 283.7906329 & -30.4768056  &   1.3476492 & 2.11223  & 17.73 & 17.24 & 16.60 & 0.01 & 1.26 & 0.94 & 0.56 & 0.675  & 0.790 & 0.532 & 0.662 &	 \\  
   02 &  0	& 283.7621242 & -30.4543426  &   1.0945410 & 2.65622  & 17.73 & 17.34 & 16.72 & 0.02 & 1.04 & 0.89 & 0.57 & 0.424  & 0.759 & 0.411 & 0.635 &	 \\  
   03 &  1	& 283.7591376 & -30.4295744  &   0.5726301 & 9.60328  & 18.70 & 18.19 & 17.52 & 0.03 & 1.09 & 0.76 & 0.55 & 0.610  & 0.745 & 0.536 & 0.687 &   \\  
   04 &  1	& 283.7454891 & -30.3931179  &   0.4832307 & 9.75278  & 18.57 & 18.32 & 17.64 & 0.03 & 1.41 & 0.77 & 0.60 & 0.590  & 0.717 & 0.338 & 0.693 &   Bl\\  
   05 &  1	& 283.7218320 & -30.4670001  &   0.5790976 & 9.70703  & 18.66 & 18.11 & 17.47 & 0.02 & 1.23 & 0.95 & 0.66 & 0.583  & 0.754 & 0.562 & 0.664 &	 \\  
   06 &  1	& 283.8307177 & -30.5287519  &   0.5384199 & 3.51936  & 18.41 & 18.03 & 17.43 & 0.04 & 1.17 & 0.71 & 0.68 & 0.596  & 0.583 & 0.420 & 0.610 &   Bl\\  
   07 &  1	& 283.7810458 & -30.5232817  &   0.5944093 & 4.57644  & 18.73 & 18.21 & 17.54 & 0.01 & 0.86 & 0.74 & 0.47 & 0.574  & 0.762 & 0.537 & 0.681 &	 \\  
   12 &  2	& 283.6919599 & -30.5478954  &   0.3226390 & 9.30830  & 17.14 & 16.71 & 16.15 & 0.03 & 0.48 & 0.34 & 0.22 & -----  & ----- & 0.435 & 0.569 &   F \\  
   14 &  1	& 283.8410760 & -30.4206737  &   0.4807214 & 3.04538  & ----- & ----- & ----- & ---- & ---- & ---- & ---- & -----  & ----- & ----- & ----- &	 \\  
   15 &  1	& 283.8053859 & -30.4962929  &   0.5868540 & 3.54588  & 18.69 & 18.21 & 17.53 & 0.02 & 0.95 & 0.93 & 0.61 & 0.493  & 0.800 & 0.481 & 0.704 &   \\  
   28 &  1	& 283.7859926 & -30.4341522  &   0.5088183 & 3.83630  & 18.48 & 18.08 & 17.51 & 0.05 & 1.40 & 1.18 & 0.78 & 0.518  & 0.734 & 0.432 & 0.614 &  Bl,Sgr \\  
   29 &  1	& 283.7226484 & -30.4924474  &   0.5897109 & 2.90001  & 18.77 & 18.25 & 17.52 & 0.02 & 0.92 & 0.72 & 0.41 & 0.575  & 0.801 & 0.532 & 0.742 &	 \\
   30 &  1	& 283.7645154 & -30.4570863  &   0.5738690 & 9.65573  & 18.74 & 18.24 & 17.55 & 0.02 & 1.10 & 0.87 & 0.56 & 0.588  & 0.777 & 0.524 & 0.701 &	 \\
   31 &  1	& 283.7323449 & -30.4985297  &   0.6466228 & 3.56726  & ----- & 18.18 & 17.43 & 0.02 & ---- & 0.40 & 0.23 & -----  & 0.802 & ----- & 0.746 &	 \\
   32 &  1	& 283.7063948 & -30.4623137  &   0.5191334 & 3.85942  & 18.49 & 18.08 & 17.46 & 0.07 & 1.27 & 0.90 & 0.72 & 0.517  & 0.772 & 0.443 & 0.644 &  Bl \\
   33 &  1	& 283.7890960 & -30.5073081  &   0.4930339 & 3.51936  & 18.75 & 18.13 & 17.59 & 0.06 & 1.12 & 0.77 & 0.74 & 0.647  & 0.604 & 0.633 & 0.560 &  Bl \\
   34 &  1	& 283.7473555 & -30.5219780  &   0.5034071 & 3.83630  & 18.68 & 18.16 & 17.61 & 0.03 & 1.38 & 1.14 & 0.70 & 0.527  & 0.767 & 0.524 & 0.604 &	 \\
   35 &  1	& 283.7377861 & -30.4650676  &   0.5267239 & 9.78589  & 18.66 & 18.20 & 17.59 & 0.03 & 1.41 & 1.09 & 0.74 & 0.587  & 0.734 & 0.500 & 0.639 &	 \\
   36 &  1	& 283.8053328 & -30.4637899  &   0.5990345 & 9.66790  & 18.70 & 18.16 & 17.47 & 0.02 & 0.80 & 0.60 & 0.41 & 0.619  & 0.757 & 0.551 & 0.701 &	 \\
   37 &  1	& 283.7784287 & -30.4907654  &   0.6274528 & 3.77083  & 18.74 & 18.21 & 17.49 & 0.01 & 0.62 & 0.48 & 0.23 & 0.566  & 0.794 & 0.529 & 0.727 &	 \\
   38 &  1	& 283.7424905 & -30.4697269  &   0.6111295 & 9.64313  & 18.58 & 18.04 & 17.30 & 0.02 & 0.60 & 0.43 & 0.28 & 0.565  & 0.788 & 0.542 & 0.743 &	 \\
   39 &  1	& 283.7323257 & -30.4975579  &   0.5995711 & 3.77083  & ----- & 18.19 & 17.53 & 0.02 & ---- & 0.76 & 0.48 & -----  & 0.755 & ----- & 0.680 &	 \\
   40 &  1	& 283.7483469 & -30.5102175  &   0.5863672 & 2.66921  & 18.74 & ----- & 17.51 & ---- & 0.90 & ---- & 0.50 & -----  & ----- & ----- & ----- &	 \\
   41 &  1	& 283.8052158 & -30.4654648  &   0.6176814 & 3.68530  & 18.77 & 18.15 & 17.43 & 0.05 & 0.97 & 1.04 & 0.61 & 0.594  & 0.870 & 0.613 & 0.752 &   Ev\\
   42 &  2	& 283.7866761 & -30.4629200  &   0.3264348 & 9.07901  & ----- & 18.19 & 17.65 & 0.02 & ---- & 0.49 & 0.31 & -----  & ----- & ----- & 0.553 &	 \\
   43 &  1	& 283.7150492 & -30.4659484  &   0.5928533 & 4.64542  & 18.73 & 18.20 & 17.46 & 0.02 & 0.87 & 0.80 & 0.57 & 0.587  & 0.830 & 0.542 & 0.757 &   Bl\\ 
   44 &  1	& 283.7688542 & -30.5014846  &   0.6207049 & 9.49051  & 18.77 & 18.25 & 17.52 & 0.02 & 0.61 & 0.39 & 0.27 & 0.590  & 0.770 & 0.538 & 0.732 &	 \\
   45 &  1	& 283.8036281 & -30.5080007  &   0.4888223 & 3.47103  & 18.62 & 18.24 & 17.63 & 0.03 & 1.32 & 1.02 & 0.62 & 0.530  & 0.752 & 0.427 & 0.640 &	 \\
   46 &  1	& 283.7530186 & -30.4903405  &   0.6047506 & 4.52668  & 18.65 & 18.08 & ----- & 0.02 & 0.64 & 0.56 & ---- & 0.603  & ----- & 0.581 & ----- &   Sgr\\
   47 &  1	& 283.7542761 & -30.4532285  &   0.5069847 & 2.88684  & 18.62 & 18.22 & 17.61 & 0.02 & 0.44 & 0.70 & 0.41 & 0.332  & 0.670 & 0.397 & 0.618 &   Bl \\
\hline																			    
\end{tabular}																		    
\end{minipage} 
\end{table}
\end{landscape}

\section{Properties of the RR Lyrae stars}\label{s:prop} 

The sample of variables detected in our survey is mainly constituted by RRL 
stars, as expected from an old metal-poor globular cluster (GC). 
In this section we discuss the mebership of the various RRL, attempting to 
define a sample of M54 RRL as clean as possible from interlopers and/or stars 
with uncertain or possibly problematic data. Then, we will use the cleaned 
sample to analyze the pulsational properties of the cluster RRL.

\subsection{The  distribution in colour and magnitude}\label{s:cmd}

\begin{figure}
 \includegraphics[width=8.7cm]{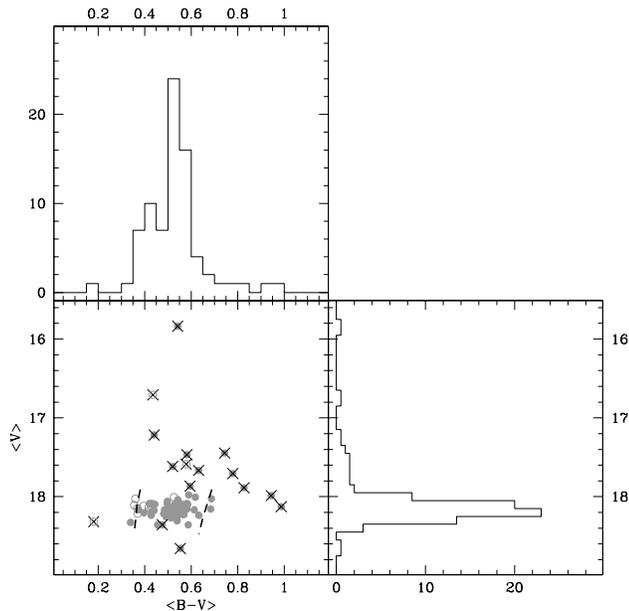}
\caption{Colour-magnitude distribution of RR Lyrae stars in the V-(B-V) plane 
(bottom-left panel). RR0 and RR1 variables are marked with filled and open grey
circles, respectively. Stars rejected from the M54 sample (see Sect. \ref{s:cmd}) are
marked with black crosses. The two dashed lines indicate the edges of the
instability strip (from Bono et al. 1995). The $<B-V>$ colour and $<V>$
magnitude distributions are shown in the upper-left and bottom-right panels,
respectively.}
\label{cmdrr}
\end{figure}

\begin{figure}
 \includegraphics[width=8.7cm]{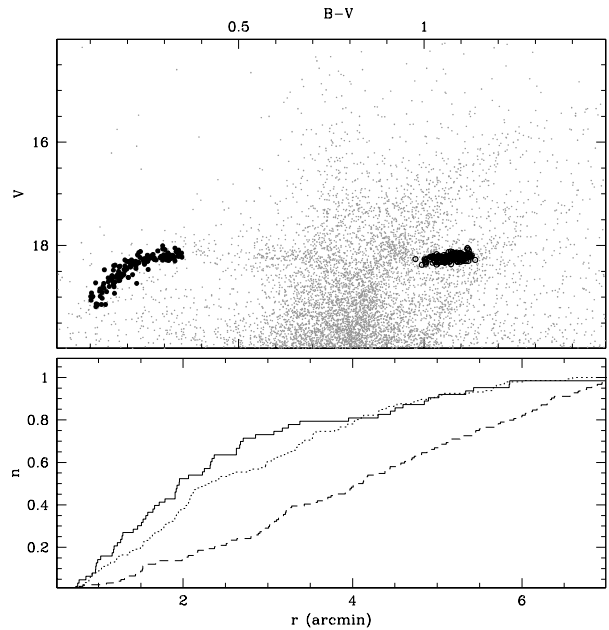}
\caption{The cumulative radial distribution of the RRL of M54 (solid line) compared 
with those of the BHB (dotted line) and red clump (dashed line) stars. The BHB and 
red clump stars are highlighted in the CMD (upper panel), for
convenience.}
\label{f:rad}
\end{figure}

We show in Fig. \ref{cmdrr} the distribution in magnitude and colour of the RRL 
stars for which we obtained calibrated photometry. 
The magnitude histogram in the lower-right panel shows that the distribution is 
dominated by a strong and relatively narrow peak around $V\simeq 18.2$, clearly 
associated to the cluster population. The bulk of this population is well 
bracketed between the edges of the instability strip predicted by  theoretical 
models (Bono, Caputo \& Marconi 1995) for a 
0.65 $M_{\odot}$ RRL star with Z=0.0004, overplotted to the CMD of Fig. 
\ref{cmdrr}. Here we assume $(m-M)_{0}=17.13$ and $E(B-V)=0.16$, according to 
the results of Sect.~\ref{s:dist}, below.

There are several stars that lie outside the M54 peak, both in magnitude and colour. 
There are 11 variables (9 RR0 and 2 RR1) having $\langle V\rangle<17.9$, 
clearly brighter than the bulk of the RRL population, i.e., in order of increasing 
magnitude, V119, V12, V128, V132, V93, V142, V65, V121, V75, V92, and V69. 
Assuming $M_V(RR)=0.6$ and $E(B-V)=0.16$ we can obtain rough 
estimates of the distances of these stars. V119 has $D\sim 9$ kpc, thus may be 
possibly located in the Galactic Bulge. The others range between $D\sim13$ kpc 
and $D\sim 23$ kpc; while some of these are likely interlopers from the 
Galactic thick disc and/or halo, the faintest ones are more likely members of 
M54 (or Sgr dSph) that appear over-luminous because of blending with other 
sources and/or some unidentified problem in the photometry; in fact four of 
them (V92, V128, V132 and V142) lie in the most crowded region of our images, 
within $1\arcmin$ from the cluster center. 

V55 (the faintest of the outliers, $D\sim 33$ kpc) is likely a genuine Galactic 
star in the background. Finally there are five stars that have at least one 
colour clearly outside the RRL window (V75, V123, V124, V133, V143; only V75 
appears anomalous in both colours): this is likely due to a problem in the 
DoPHOT photometry of the reference frame in one of the three bands. 
V123, that appears rather normal from Fig.~\ref{cmdrr} has a very red V-I.

To keep our final sample of M54 RRL as clean and safe as possible, all the 
stars discussed above have been excluded from the analysis of the pulsational 
properties, in the following.
The RRL of this clean sample (55 RR0 and 9 RR1)  have  
$17.9<V<18.4$ mag and $0.3 < B-V < 0.7$.
In Fig.~\ref{f:rad} the cumulative radial distribution of the RRL in the clean 
sample is compared with those of the Blue HB stars (BHB; expected to be 
dominated by M54 population) and red clump stars.
As expected, both the RRL and the BHB stars are significantly more 
concentrated than the red-clump stars. 
This indicates that the contamination by Galactic and/or Sgr variables should 
be minimal (see also Monaco et al. 2003). 

\begin{figure}
 \includegraphics[width=8.7cm]{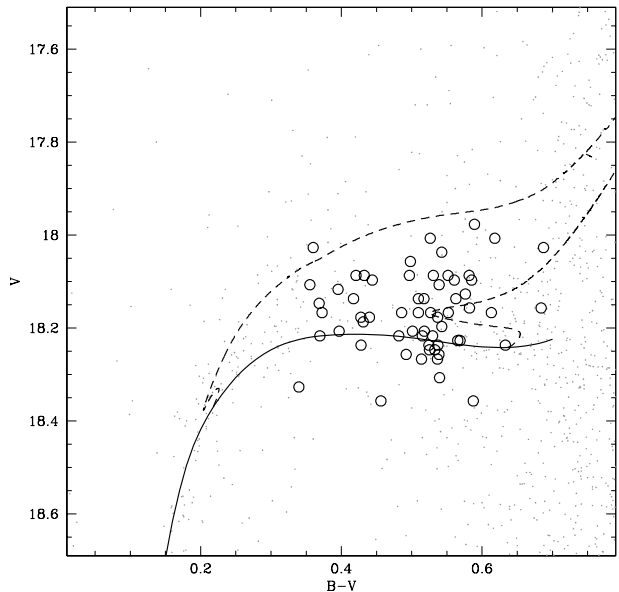}
\caption{Colour-magnitude distribution of the HB area in the 
V-(B-V) plane. RR Lyrae stars are shown as open circles. 
The ZAHB model (solid line) for metallicity $Z=0.0006$ and helium abundance $Y=0.23$, 
and the evolutionary tracks (dashed lines) for the two stellar masses M=0.65 and 
0.70~$M_{\odot}$ (from Cariulo et al. 2004) are overplotted.}
\label{theo}
\end{figure}

In Fig. \ref{theo} the Zero Age Horizontal Branch (ZAHB) model (from Cariulo,
Degl'Innocenti \& Castellani 2004) for a population
with metallicity $Z=0.0006$ and helium abundance $Y=0.23$ is overplotted to the
$V-(B-V)$ CMD of M54. 
Two evolutionary tracks corresponding to mass values of 
$M=0.65~M_{\odot}$ and $M=0.70~M_{\odot}$ are overplotted.
Absolute magnitudes have been converted to apparent ones adopting the reddening 
and distance modulus derived in Sect. \ref{s:redd} and \ref{s:dist}.
The ZAHB model fits quite well the lower envelope of the RRL 
distribution. 
Three stars are about 0.2 mag fainter than the average, 
namely V4, V118 and V130. The variable V4 is affected by the Blazhko effect and 
we have likely missed the maximum, as suggested by the unusually low V amplitude 
(0.71 mag) for its period (0.48 d). The other two stars have normal well defined light 
curves, and no metallicity estimates are available to check the origin of 
their unusual magnitude.

\subsection{Pulsational properties}\label{s:pac}

If we consider the entire sample of RRL, including also those for which only internal 
calibrated photometry could be obtained (but excluding those flagged as non-member or 
problematic in the previous section), then we have 95 RR0 stars with 
$<P_{ab}>$ = 0.584d and 33 RR1 stars with $<P_{c}>$ = 0.351d. 
For comparison, the average values for Oosterhoff I type clusters are 
$<P_{ab}>$ = 0.559d and $<P_{c}>$ = 0.326d, while for Oosterhoff II
type clusters they are $<P_{ab}>$ = 0.659d and $<P_{c}>$ = 0.368d (Smith 1995).  
The fraction of RR1 over the entire sample of RRL is 0.26,  
which is slightly higher than the typical fraction detected in Oosterhoff I 
clusters (0.22) and significantly smaller than the typical value (0.48) in Oosterhoff 
II clusters  (Clement et al. 2001). This indicates that the entire sample 
of RRL stars detected in the M54 region has intermediate characteristics 
between the classical Oosterhoff types I and II, being much closer to Oosterhoff I. 
If we consider only the subset of stars that are most likely M54 members  
according to luminosity and metallicity criteria, as we discuss in Sect. \ref{s:met}, 
we find $<P_{ab}>$ = 0.600 d from 49
RR0 stars and $<P_{c}>$ = 0.347 d from 7 RR1 stars

\begin{figure}
 \includegraphics[width=8.7cm]{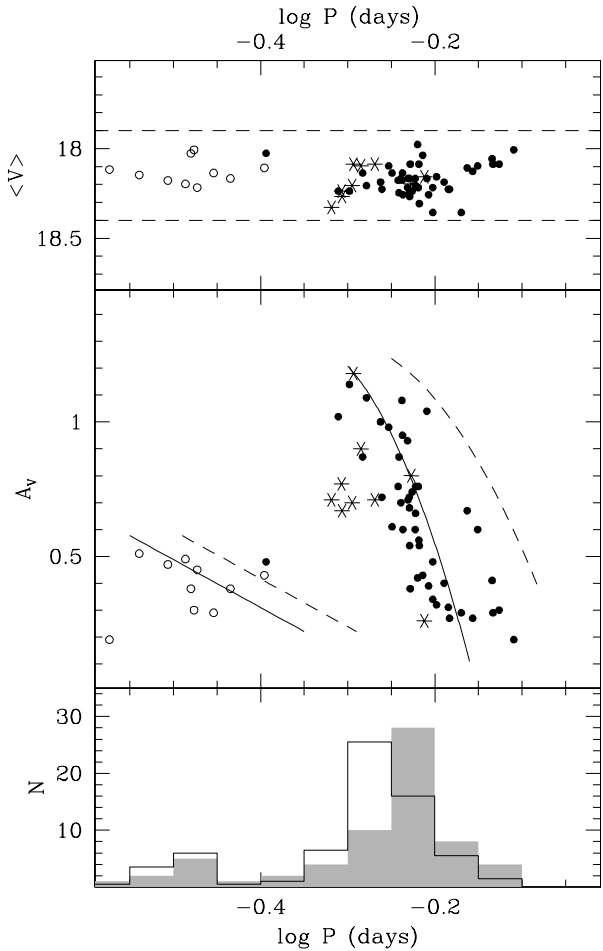}
\caption{Upper panel: Period - $<V>$ distribution, cleaned of field contaminants, 
possible blends and photometric errors. Fundamental (RR0) and first overtone (RR1) pulsators are marked with filled 
and open circles, respectively. Blazhko stars are indicated as asterisks. The
limiting magnitude of our selection criterion are marked with dashed lines. 
Middle panel: Period - V amplitude diagram for the RRL of M54. 
The symbols are the same
as the upper panel. The mean loci of the unevolved RRL in M3 (solid line) 
and M92 (dashed line) are overplotted (from Cacciari et al. 2005).
Lower panel: Period distributions for the RRL of M54 (grey histogram) and 
M3 (open histogram).} 
\label{pa}
\end{figure}

In the middle panel of Fig. \ref{pa} the period-V amplitude diagram for the sample of RRL is shown.
The mean loci of the unevolved RR0 and RR1 variables in the GGCs M3 
(Oosterhoff I) and M92 (Oosterhoff II) are overplotted (from Cacciari, Corwin \& Carney 2005). 
Taken at face value, the majority of the RR0 fall around the ridge line of M3 
slightly on the side of smaller amplitudes or shorter periods. 
Some of these stars could well be 
unrecognised Blazhko variables observed during the low amplitude phase of the 
Blazhko modulation. Indeed, Blazhko effect has been 
clearly detected for 9 RRLs of our sample. According to previous studies (Smith 1995; Corwin \& Carney 2001; 
Cacciari et al. 2005) as much as 30\% of the total RRL population 
in several clusters is affected by Blazhko modulation. If the M54 population 
shares the same behaviour, then at least twice as many Blazhko RRL 
variables, whose anomaly could 
not be detected because of the short time coverage of our observations, are expected to exist in the studied sample . 
Within these uncertainties, the M54 data appear to be in good agreement with  
the mean relations derived for M3, albeit with a somewhat larger scatter (that may be due 
to the intrinsic metallicity spread of the cluster, see C10).  
In the bottom panel of Fig. \ref{pa} the period distribution of
the RRL of M54 is compared with that of M3. 

The apparent contadiction between the location in the P-Av diagram (slightly
on the side of shorter periods) and the (longer) mean period of the M54 RRab
stars with respect to M3 is an optical effect of the different
amplitude distributions: while 44\% of M3 RRab have amplitudes $A_{V}>1~mag$,
this fraction is only 11\% in M54. At least part of this difference might be
due to the presence of a significant fraction of unrecognised Blazhko variables.
As for the period distributions, which are not affected by the Blazhko stars,
it is evident that the M54 and M3 RRL period distributions are different: the M3
distribution is skewed towards the short period end whereas in M54 it is wider and
more symmetrical (bottom panel of Fig.9). This accounts for a longer mean RRab period
in M54 ($\sim$0.60d) than in M3 ($\sim$0.56d). 

One can also note a group of 8 stars with longer periods that seem
to stand out from the main distribution . These stars are V41, V48, V76, V83,
V120, V125, V137 and V140. This location in the period-amplitude diagram
can be due to evolution off the ZAHB  or to lower metallicity. In both cases
the star is expected to be somewhat overluminous.
Two of them (V125 and V137) are indeed brighter than the average by more than 0.1
mag and might be evolved, for the other 6 stars the luminosity is
quite normal.  The metallicity was estimated for 4 of these stars
(V41, V48, V76 and V83) and falls within about 0.2 dex of the mean value
except for V83 that is very metal-poor (see Sect. \ref{s:met}).

In conclusion M54 RRL
show Oosterhoff I characteristics as far as the population ratio 
of fundamental/first overtone pulsators is concerned, while display average periods 
quite intermediate between the two types, mainly due to the presence of
a significant number of long period RRL.  

\section{Metallicity estimates from Fourier analysis}\label{s:met} 

It was shown that appropriate combinations of terms of a Fourier 
representation of RRL light curves, along with the periods, correlate with 
intrinsic parameters of the stars, such as metallicity, mass, luminosity and 
colours (Simon 1988; Clement, Jankulak \& Simon 1992; Jurcsik \& Kovacs 1996; Kovacs \& Walker 2001; 
Morgan, Wahl \& Wieckhorst 2007). Sandage (2004) interpreted the metallicity
dependence  of the Fourier components of RRL light curves as equivalent to 
the Oosterhoff-Arp-Preston period-metallicity effect. This can be found 
in both fundamental and first overtone pulsators, that we consider here 
separately. 

However, in several cases (Jurcsik \& Kovacs 1996; Pritzl et al. 2001;
Nemec 2004; Cacciari et al 2005; Stetson, Catelan \& Smith 2005) the coefficients as well as
the range of validity of such relations have been discussed, often leading to
different and conflicting results. Therefore, [Fe/H] estimates derived from
the Fourier parameter $\phi_{31}$ should be used with caution, especially in the high
metallicity regime. In the following we use these metallicites mainly to clean the
sample of M54 RRL stars from possible members of Sgr, that would lie at the same
distance but may stand out as significant outliers in the metallicity distribution.

To this purpose, we have decomposed the V light curves of all RRL variables 
of our sample in Fourier series of cosines. 

\subsection{RR1}\label{s:met_rrc} 

\begin{table}
\label{t:ffec}
 \centering
 \begin{minipage}{85mm}
  \caption{Fourier parameter $ \phi_{31}$ (cosine series) and [Fe/H]  
  for the RR1  variable stars. In column 5 the Notes indicate: * = excluded for its high/low magnitude (see
  Sect.\ref{s:cmd}); ** =
  excluded from the analysis for its high/low metallicity (see Sect.
  \ref{s:met_rrab}).}
  \begin{tabular}{@{}lcccl@{}}
  \hline
   ID  & Period  & $\phi_{31}$ & $[Fe/H]_{ZW}$ & Note \\
       & (d)     &             &        & \\
 \hline \\
   V12 &  0.3226390 & 4.277  &  -1.01 & * \\  
    V42 &  0.3264349 & 3.140  &  -1.73 & \\  
   V49 &  0.3312773 & 3.867  &  -1.40 & \\  
   V56 &  0.3676695 & 3.741  &  -1.81 &   \\  
   V74 &  0.2889211 & 4.267  &  -0.49  & ** \\    
   V78 &  0.4019678 & 4.473  &  -1.72  &    \\ 
   V97 &  0.3367274 & 3.653  &  -1.58  &    \\  
   V126 & 0.3516532 & 3.810  &  -1.65  &    \\  
   V133 & 0.3648148 & 3.782  &  -1.77  &    \\  
   V138 & 0.3116047 & 3.059  &  -1.62  &   \\  
   V141 & 0.3339446 & 4.327  &  -1.13  &  **   \\    
   V142 & 0.2978301 & 2.333  &  -1.79  &  *\\ 
   V148 & 0.2670186 & 4.149  &  -0.19  &  **\\
\hline
\end{tabular}
\end{minipage}
\end{table}

The use of Fourier coefficients to characterise RR1 variables has been 
less extensive than that of the RR0 stars. The most recent critical analysis of 
this method has been presented by Morgan et al. (2007). 
We refer to this work for an extended discussion and references.
According to these authors the metallicity of the RR1 stars can be estimated
using the following relation:

$$[Fe/H]_{ZW} = 52.466P^2 - 30.075P + 0.131\phi_{31}^2 +0.982\phi_{31}$$  
$$- 4.198P\phi_{31} +2.424~~~~~~~~~~~~~~~$$

\noindent where [Fe/H]$_{ZW}$ is expressed in the 
Zinn \& West (1984) metallicity scale, 
and the coefficient $\phi_{31}$ comes from Fourier light curve decomposition in series of 
cosines. The standard deviation of this relation, obtained from 106 stars, is 0.145 dex. 
The standard deviation of this relation, obtained from 106 RRc stars in 12
globular clusters, is 0.145 dex. The [Fe/H] values of the calibrating clusters are between 
approximately -1  and -2, so this relation likely gives poorer results at very high or very
low metallicities.

Our sample contains 13 RR1 stars for which we have estimated the Fourier parameter 
$ \phi_{31}$ and hence the metallicity with the above relation. 
The results are listed in Table 2. 
Two stars (V12 and V142) have been assigned to the 
Galactic field population based on their unusually bright V magnitude
($V<17.9$).
Three more stars (V74, V141 and V148) have rather high metallicities 
suggesting that they may belong to the Sgr dSph population.
The average metallicity of the 8 stars that are most likely M54 members is 
[Fe/H]= -1.66, with a standard deviation of 0.13 dex. 

\subsection{RR0}
\label{s:met_rrab} 
 
\begin{figure}
 \includegraphics[width=8.6cm]{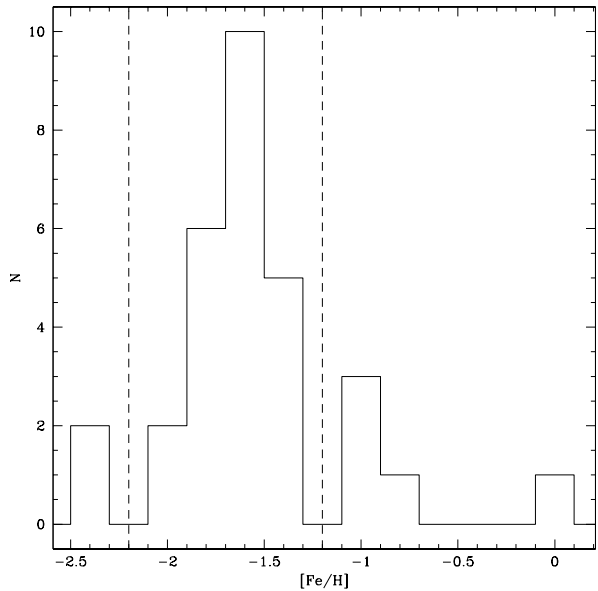}
\caption{Metallicity distribution of the 30 RR0 variables with $D_{m}\le5$ and
$17.9<V<18.4$. 
The range of the adopted selection for the {\it bona fide} sample of M54 are
marked with dashed lines.}
\label{f:met}
\end{figure}
 
From the analysis of  V light curves of 272 RR0 stars, Jurcsik \& Kovacs (1996)
and Jurcsik (1998) found a 
tight correlation (r.m.s. error of the fit 0.14 dex) 
between the  metallicity of the variables and their $\phi_{31}$ (sine series) Fourier 
coefficents: $[Fe/H]=-5.038 - 5.394~P + 1.345~\phi_{31}$. 
Based on the Layden (1994) [Fe/H] values of a large number
of field RRab stars, Sandage (2004) recalibrated the above relation and obtained
a lower zero-point by about -0.25 dex that is consistent with the Layden
metallicity scale. This relation, that we use in the present analysis, is:

$$[Fe/H]_L = -5.492 - 5.666~P + 1.413~\phi_{31}$$

\noindent where the r.m.s. error of the fit remains $\sim$ 0.14 dex.  
This new formulation gives values of metallicity that are in good agreement
with those obtained from spectroscopy or other methods, as we comment later.
Since we are also interested in differential estimates, to separate the M54
from the (Milky Way and Sgr) field populations, this method appears sufficiently
adequate to identify at least the most obvious outliers.
We corrected the values of our cosine series $\phi_{31}$ by +$\pi$ to report 
them to sine series decomposition, and applied 
the above relation to calculate the metallicity. 
The $D_{m}$ parameter (defined by Jurcsik \& Kovacs 1996) 
represents a quality test on the regularity of the shape of the light 
curves, and hence on the reliability of the derived physical parameters.  
We considered as sufficiently reliable only stars with $D_{m} \le 5$.   
The values of $\phi_{31}$ and [Fe/H]$_L$ for these 38 RR0 stars are listed 
in Table 3. 
Blazhko stars should be excluded, unless observed at or near 
maximum Blazhko modulation amplitude with a good degree of confidence.  
In fact, at small-amplitude Blazhko phase the light curves are more likely 
distorted (even if Dm $\le$5) and tend to overestimate the metallicity, 
whereas at large-amplitude Blazhko phase they are quite similar to regular 
pulsators (see Cacciari et al. 2005). 
Following these considerations, only V33 has been excluded from the sample.
On the other hand, photometric blends or stars affected by calibration errors 
can still produce good metallicity estimates, as only the zeropoint of the 
photometric scale is incorrect.  
Seven variables with a reliable metallicity estimate have been associated to the 
foreground Galactic population because of their bright magnitudes. 

The distribution of the remaining 30 RR0 variables with reliable metallicity estimates 
is shown in Fig. \ref{f:met}.  
As can be noted, the distribution has a prominent peak at $[Fe/H]\sim-1.6$ 
and the majority of RR0 variables have metallicities that range between 
$-2.2<[Fe/H]<-1.2$. A group of four variables (V28, V46, V63 and V123) form a 
secondary peak at $[Fe/H]\simeq -1.0$. One variable (V139) has a solar-like
metallicity. Finally, two variables (V50 and V83) present a very low metallicity 
($[Fe/H]<-2.2$). 
The membership of these 7 stars with metallicities outside the range
$-2.2<[Fe/H]<-1.2$ is not clear cut.
The extreme metallicity of V139, V50 and V83 strongly suggests a membership to Sgr dSph, 
whose metallicity distribution tails clearly reach values from  
$[Fe/H]\sim-2$ up to nearly solar (B08; Monaco et al 2005b; Bonifacio et al.
2006). 
In particular Cseresnjes (2001) found a very wide metallicity distribution also for 
the RRL of Sgr, with a mean of [Fe/H]=-1.6, a dispersion of 
$\sim 0.5$ dex, and a minor but significant population at $[Fe/H]\la$-2.0 dex.
The membership of the four stars at $[Fe/H]\simeq -1.0$ is harder to establish, 
especially considering that M54 itself has been found (from spectroscopic analyses) 
to display a spread of $\sim 0.19$ dex (Da Costa \& Armandroff 1995;
B08; C10). To be conservative we decided to exclude also these 
four stars from our final sample\footnote{It is worth noting that although this 
selection criterion is consistent with the metallicity distributions
of about 100 RGB stars of M54 and surrounding Sgr field found by C10, 
the metallicity distributions of M54 and Sgr do show some overlap.
Therefore, our bona fide M54 sample may still contain some (very few) Sgr field
stars that might be distinguished by only the Na/O anticorrelation cluster
signature.}.

The remaining 23 RR0 variables that are likely members of M54 have a mean metallicity  
[Fe/H]= -1.65 $\pm$ 0.18 dex. 
The mean metallicity of the M54 RR0 stars is in excellent agreement with the 
mean value found for the RR1 in Sect. \ref{s:met_rrc}. 
So we can assume that the RR1 and RR0 metallicity estimates can be dealt with 
jointly, and  the mean value of the combined sample is   
[Fe/H]= -1.65, with a standard deviation of  0.16 dex, 
in good agreement with the spectroscopic estimates 
by Da Costa \& Armandroff (1995), Brown et al. (1999), B08 and C10. The 
uncertainties on individual measures, on the reliability of the
$\phi_{31}-[Fe/H]$ relations and the conservative 
selection described above prevent any conclusion about the intrinsic 
metallicity spread.

By rejecting those RRL with measured metallicities outside the range
$-2.2<[Fe/H]<-1.2$
, in addition to the luminosity criteria described in Sect.
\ref{s:cmd}, 
we end up with a sample of 49 RR0 and 7 RR1 variables, that will be used in 
the following sections to 
estimate the distance modulus and the reddening of M54. 

In the following, we 
will refer to these stars as the {\it bona-fide} M54 sample.

\begin{table}\label{t:ffe}
 \centering
 \begin{minipage}{85mm}
  \caption{Fourier parameter $ \phi_{31}$ (cosine series) and [Fe/H]  
  values for the RR0  variable stars with Dm $<$5 in the M54 field of view. 
  In column 5 the Notes indicate: * = excluded for its high/low magnitude (see
  Sect.\ref{s:cmd}); ** =
  excluded from the analysis for its high/low metallicity (see Sect.
  \ref{s:met_rrab}).}
  \begin{tabular}{@{}lcccl@{}}
  \hline
   ID  & Period  & $\phi_{31}$ & [Fe/H] & Note \\
       & (d)     &             & L84    & \\
 \hline \\
   V3  &  0.5726301 &  1.880 & -1.64   &	\\  
   V5  &  0.5790976 &  1.702 & -1.93   &	\\  
   V7  &  0.5944093 &  1.956 & -1.65   &	\\  
   V15 &  0.5868540 &  1.906 & -1.69   &	\\  
   V28 &  0.5088183 &  2.030 & -1.07   & Blazhko, ** \\    
   V29 &  0.5897109 &  2.115 & -1.41   &	\\ 
   V30 &  0.5738690 &  1.994 & -1.49   &	\\  
   V32 &  0.5191334 &  1.521 & -1.85   & Blazhko \\  
   V33 &  0.4930339 &  1.894 & -0.66:  & Blazhko, low-ampl. \\  
   V34 &  0.5034071 &  1.786 & -1.38   &	\\  
   V35 &  0.5267239 &  1.573 & -1.81  &   \\ 
   V41 &  0.6176814 &  2.202 & -1.44  &   \\
   V43 &  0.5928533 &  2.020 & -1.56  & Blazhko   \\
   V45 &  0.4888223 &  1.404 & -1.84  &   \\
   V46 &  0.6047506 &  2.629 & -0.76  & **  \\
   V48 &  0.6870139 &  2.262 & -1.75  &   \\
   V50 &  0.5636133 &  1.258 & -2.47  & **  \\
   V55 &  0.4260544 &  2.154 & -0.42  &  * \\
   V58 &  0.5486673 &  1.797 & -1.62  &   \\
   V61 &  0.6039009 &  2.165 & -1.42  &   \\
   V63 &  0.5901396 &  2.361 & -1.06  &  ** \\
   V65 &  0.5748202 &  2.045 & -1.42  &  * \\
   V67 &  0.5900106 &  1.896 & -1.72  &   \\
   V69 &  0.6797878 &  2.006 & -2.07  &  * \\
   V75 &  0.5846341 &  1.907 & -1.67  &  * \\
   V76 &  0.7064914 &  2.378 & -1.70  &   \\
   V77 &  0.5769039 &  1.867 & -1.68  &   \\
   V82 &  0.5878700 &  2.020 & -1.53  &   \\
   V83 &  0.5783485 &  1.398 & -2.35  &  ** \\
   V85 &  0.5213111 &  1.381 & -2.06  &   \\
   V92 &  0.4840892 &  1.557 & -1.60  &  *\\
   V93 &  0.5585378 &  1.780 & -1.70  &  *\\
   V96 &  0.5585253 &  1.883 & -1.56  &   \\
  V119 &  0.6366165 &  2.356 & -1.33  &  *\\
  V123 &  0.5141767 &  2.112 & -0.98  &  ** \\
  V124 &  0.5822650 &  1.812 & -1.79  &   \\
  V129 &  0.6027318 &  2.060 & -1.56  &   \\
  V139 &  0.4039767 &  2.337 & -0.04  &  ** \\
\hline
\end{tabular}
\end{minipage}
\end{table}

\section{Reddening and Distance}\label{s:redd}

Our large sample of RRL allows us to determine the reddening  to M54 with  
good accuracy.
We consider two approaches to estimate the reddening, using the 
colours of the RR0 variables at minimum light (i.e. averaged over
the phase interval 0.5-0.8):
\begin{itemize}
\item
Based on a recalibration of the original relation by Sturch (1966), Walker (1998) 
proposed a relation linking $(B-V)_{min}$ and reddening $E(B-V)$: 
$$E(B-V)=(B-V)_{min}-0.24~P-0.056~[Fe/H]-0.356$$
Applying this relation to the sample of {\it bona-fide} RR0 variables 
of M54 we find $E(B-V)=0.17\pm0.02$ if we adopt an average 
metallicity $[Fe/H]=-1.65$ (see Sect. \ref{s:met}) and 
$E(B-V)=0.16\pm0.06$ if we adopt for each star the metallicity estimated 
through the $\phi_{31}-[Fe/H]$ relation defined in Sect. \ref{s:met_rrab}

\item
From the analysis of a sample of field RR0 stars Mateo et al. (1995) 
proposed a constant value of $(V-I)_{0,min}=0.58$  for RR0 stars, 
independent on metallicity and period. 
By using our sample of {\it bona-fide} M54 RR0 stars 
we find $<(V-I)_{min}>=0.78\pm0.03$ which translates into 
$E(V-I)=0.20\pm0.03$, and hence $E(B-V)=0.16\pm0.03$ 
using the relation E(B-V)=0.8 E(V-I)(Dean, Warren \& Cousin 1978).

\end{itemize}

Summarizing the above results, we estimate an average value of 
$E(B-V)=0.16\pm0.02$. 
This estimate is in very good agreement with that 
provided by the extinction map by Schlegel, Finkbeiner \& Davis (1998,  
$E(B-V)=0.15\pm0.02$), and with the results by LS2000  from a similar 
analysis of RRL stars ($E(V-I)=0.18\pm0.02$). 

\subsection{Distance determination}\label{s:dist}
The distance to M54 has been estimated from the mean characteristics
of the {\it bona fide} sample using the absolute visual magnitude M$_V$ that can be derived 
from the luminosity-metallicity relation 
$$M_{V} = 0.214~[Fe/H] + 0.885$$
(Clementini et al. 2003) where the adopted slope 0.214  
appears to be supported by the most accurate studies of field RRL stars 
in the Milky Way (Fernley et al. 1998; Chaboyer 1999), in the Large 
Magellanic Cloud (LMC; Gratton et al. 2004) and GCs in M31 (Rich et al. 2005).
The zero-point of such a relation has been set to be consistent with the mean 
dereddened V magnitude of the RR0 variables observed in the LMC by Clementini 
et al. (2003) ($<V_{0}>=19.064$) assuming a distance modulus of $(m-M)_{0}=18.50$ 
(Freedman et al. 2001).  

We use the mean metallicity  $[Fe/H]= -1.65 \pm 0.16$ dex (derived from the subset 
described in Sect. \ref{s:met}) to estimate $<M_V> = 0.53 \pm 0.05$ mag from the above 
equation, and the mean V magnitude  $<V> = 18.16 \pm 0.07$ mag (derived from the
stars that have been included in the {\it bona fide} sample). 
Therefore, assuming $A_{V}/E(B-V)=3.1$ (Savage \& Mathis 1979) and  
E(B-V)=0.16 $\pm$ 0.02 (see Sect. \ref{s:redd}) we estimate a distance modulus to M54
$(m-M)_{0}=17.13\pm0.11$, which 
translates into a distance to M54 of $d=26.7 \pm 1.1$ kpc. 
As a sanity check, we estimate the distance modulus also adopting the
individual Fourier metallicities listed in Tables 2 and 3
for the smaller sample of 30+7 RR0+RR1 variables,
and we obtain $(m-M)_{0}=17.11\pm0.13$.
 
The above estimates are in excellent agreement with those derived by Monaco et al. (2004; $(m-M)_{0}=17.10\pm0.15$), 
and by LS2000, once their estimate is corrected using the most recent metallicity estimates ($(m-M)_{0}=17.13\pm0.09$; see Monaco et al. 2004). 
A significantly smaller distance 
($(m-M)_{0}=16.97\pm0.07$) has been derived by Kunder \& Chaboyer (2009) on the
basis of the analysis of differential distance modulus of a sample of Sgr RRL 
with respect to a sample of RRL belonging to the Galactic bulge.
These authors adopt a different luminosity-metallicity relation (from Bono,
Caputo \& Di Criscienzo 2007), and their sample is affected 
by larger amount of extinction (E(B-V)=0.36), lying in fields at lower 
Galactic latitude, from $4\deg$ to $10\deg$ apart from M54. Moreover, the 
derived distance clearly depends on the assumed distance to the Galactic Bulge.
Finally, a lower distance in that wing of the Sgr dSph may be a real effect of 
the three dimesional orientation of the galaxy major axis. 
All these factors can contribute to account for the difference between their 
distance estimate and our results (as well as those by LS2000 and 
Monaco et al. 2004; see the latter paper for a detailed discussion and 
references on previous distance estimates). 

\section{Other variables}\label{s:other}
  
\begin{figure}
 \includegraphics[width=8.7cm]{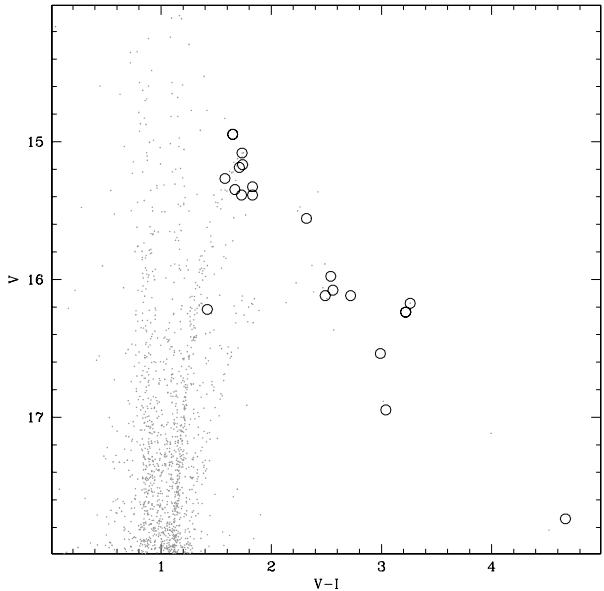}
\caption{V-(V-I) CMD of M54 and Sgr. The locations of the 18 candidates LPV are
marked with open circles.}
\label{lpv}
\end{figure}

Apart from the RRL, a number of other variables have been found in the present
survey.

Variables V1 and V2 are Cepheids already discovered by Rosino \& Nobili (1959)
and classified by LS2000 as a population II and an anomalous cepheid, respectively. 
Their magnitudes are
consistent with those predicted by the period-luminosity relation of 
Nemec, Nemec \& Lutz (1994) and are located at 3.2 and 3.5 arcmin from the
cluster center, respectively. 
On the basis of these
considerations they should be likely members of M54 (or Sgr dSph).

Among eclipsing binaries, stars V116 and V195 are Algol binaries (EA), whereas V147 
and V211 are suspected EB types. 
Five variables (namely V99, V117, V134, V135, and V144) are probably W UMa (EW). 
By applying the period-colour-V luminosity relations by Rucinski (2000) we estimated
the distance moduli of these stars. Among these 5 stars, only V134 and V144 have 
distance moduli $17.3<(m-M)_{0}<17.5$ which are compatible with that of M54
estimated in Sect. \ref{s:dist}, being probable cluster
members.  

We found 3 SX Phoenicis variables (V146, V149 and V150) with periods 
in the range $0.06~d<P<0.08~d$. Adopting the period-luminosity relation by
Poretti et al. (2008) we estimated the distances for these stars. V146 and V149 
are significantly less distant than M54 ($\Delta~(m-M)_{0} \sim 0.6-1.0$ mag) and are 
likely foreground Galactic variables. The distance modulus of V150 is compatible
(within 0.3 mag) with that of M54 and could be a cluster or Sgr member. 

Finally, we identified 18 candidate long-period variables (LPV). 
Four of them (V153, V154, V155 and V156) are new discoveries. They are located
in the CMD close to the tip of the RGB at colours $(V-I) > 1.4$
(see Fig. \ref{lpv}).
Most of them are located far from the RGB ridge line of M54, on the expected 
locus of late M giants of Sgr, being likely associated to Sgr.

Of course, the short time coverage of our observations does not allow us to 
establish their pulsation periods and their average magnitudes and colours are 
uncertain. 

\section{Discussion and Conclusions}

We presented a survey of variable stars in a 13 arcmin square field of view centred 
on the GC M54 using 
deep multi-epoch observations collected with the ESO-Danish 1.54m telescope.
From this analysis we detected 177 variables, 94 of them never observed before. 
We provide the ephemerides for the entire sample of variables 
(except for 18 long-period candidates), and the 
epoch-photometry and light curves for the large subset of variables with 
calibrated photometry. 

The period-amplitude diagram of the RRL of M54 and the relative fraction 
of RR1 variables indicate that this cluster shares some of the properties of  
Oosterhoff I clusters (as already found by LS2000), but contains a larger 
number of long-period variables. 
The average period of its RR0 and RR1 stars are indeed slightly larger 
than typical Oosterhoff I cluster values, even if significantly smaller than
Oosterhoff II. 
 
Accurate estimates of the cluster reddening 
($E(B-V)=0.16\pm0.02$) and distance modulus ($(m-M)_{0}=17.13\pm0.11$) 
were also derived, 
in excellent agreement with previous estimates in literature.

The metallicity distribution of the RRL, derived through the Fourier 
parameter $\phi_{31}$, shows a clear peak at the metallicity of M54 
($[Fe/H] \sim -1.65 \pm 0.16$ dex) and a few stars with a significantly
different metallicity (outside the range $-2.2<[Fe/H]<-1.2$), likely associated with the population of the Sgr dSph. 
The presence of a extended blue HB (reaching the instability strip) 
associated with the Sgr galaxy has been already discovered by Monaco et al. 
(2003). 
We note that the presence of RRL stars as metal-rich as $[Fe/H]\ga -0.8$ (see also Cseresjnes 2001), if 
confirmed, would indicate that Sgr was able to reach 
such relatively high metallicity at a very early epoch ($\ga 10$ Gyr ago).
Unfortunately, the small number of stars belonging to this group and the large 
uncertainties in the metallicities do not allow a firm conclusion on this issue
until spectroscopic confirmation is obtained.

\section*{Acknowledgements}
This work is largely based on the Master Thesis of by S. Colucci (2000), at the Bologna University. 
We warmly thank Christine Clement, the referee of our paper, for her helpful
comments and suggestions that have helped us improve the paper significantly.
We also thank Alfred Rosenberg for providing his photometric catalog of M54 that 
was used for the calibration of the B photometry. 
This research was partly supported by the PRIN-INAF 2007 grant CRA 1.06.10.04 assigned to the project: 
"The local route to galaxy 
formation" and by the Spanish Ministry of Science and Innovation (MICINN) under the
grant AYA2007-65090.

\clearpage

\appendix

\section{Comments on individual variables}

\begin{itemize}
 \item V4: V magnitude $\sim$ 0.2 mag fainter than the bulk M54 RRL population.
 Affected by the Blazhko effect, the maximum has likely been missed, as suggested by the unusually
 low V amplitude (0.71 mag) for its period (0.48 d).
 \item V12: we have derived a more accurate period (P=0.322639), which fits 
both our data and LS2000 data, whereas LS2000 estimate (P=0.31985) produces 
out-of-phase curves. 
The star is bright and its Fourier metallicity is [Fe/H]=-1.01, this 
suggests that it may belong to the Galactic field population. 
 \item V14: classified as RR0 with P=0.70489 by LS2000, this star shows a good 
light curve (internal photometry only) with P=0.4807214 in clear disagreement 
also with Rosino \& Nobili (1959) determination. 
 \item V28: Metal-rich ([Fe/H]=-1.07). Normal V magnitude and position in the
 Av-logP diagram. Possible Sgr member.
 \item V41: Slightly shifted in the Av-logP plane towards long periods. Normal magnitude. 
 Fourier metallicity [Fe/H]= -1.44.
 \item V46: Metal-rich ([Fe/H]=-0.76). Normal V magnitude and position in the
 Av-logP diagram. Possible Sgr member.
 \item V48: Shifted in the Av-logP plane towards long periods. Normal magnitude.
 Fourier metallicity [Fe/H]= -1.75. Possibly belonging to Sgr. 
 \item V50: Metal-poor ([Fe/H]=-2.47). Normal V magnitude and position in the
 Av-logP diagram. Possible Sgr member.
 \item V55: faint magnitude, likely background Galactic field star. This is confirmed 
by its location in the P-Av diagram, much to the short-period side of the main 
distribution for the M54 stars, also supported by the metallicity obtained from 
the Fourier parameters ([Fe/H]=-0.43). 
 \item V63: Metal-rich ([Fe/H]=-1.06). Normal V magnitude and position in the
 Av-logP diagram. Possible Sgr member.
 \item V65:  bright. Fourier metallicity [Fe/H]= -1.42 . 
Normal position in the Av-logP diagram, likely field star.
 \item V69:  bright. Fourier metallicity [Fe/H]= -2.07 . 
Shifted toward long periods in the Av-logP diagram, field star or evolved.
 \item V74: Metal-rich ([Fe/H]=-0.49). Normal V magnitude and position in the
 Av-logP diagram. Possible evolved or Sgr member.
 \item V75: unusually red colours (B-V = 0.78; V-I = 0.45). Slightly brighter than average. 
 Fourier metallicity  [Fe/H]=-1.67 dex. Normal position in the Av-logP diagram. 
 Photometric calibration error (See Sect. \ref{s:cmd}).
 \item V76: Shifted in the Av-logP plane towards long periods. Normal magnitude.
 Fourier metallicity [Fe/H]= -1.70. Possibly belonging to Sgr. 
 \item V83: Metal-poor ([Fe/H]=-2.35). Normal V magnitude. Slightly shifted in the 
 Av-logP plane towards long periods. Possible Sgr member.
 \item V92: slightly brighter than average and redder than expected from its period. 
Fourier metallicity  [Fe/H]=-1.60 dex. Located within central 1 arcmin, 
possibly a blend.  
 \item V93: bright. Fourier metallicity  [Fe/H]=-1.71 dex.  Likely field star.
 \item V118: V magnitude $\sim$ 0.2 mag fainter than the bulk M54 RRL population.
 Normal location in the Av-logP plane. 
 \item V119: very bright. Fourier metallicity [Fe/H]=-1.33 dex. Foreground 
star (Galactic population). 
 \item V120: Shifted in the Av-logP plane towards long periods. Normal magnitude.
 Possibly belonging to Sgr. 
 \item V121: bright. Likely field star.
 \item V123: unusually red V-I colour (V-I = 1.00) in spite of normal B-V
(B-V=0.47). Fourier metallicity [Fe/H]=-0.98.
Photometric calibration error (See Sect. \ref{s:cmd}).
 \item V124: unusually red B-V colour (B-V = 0.94) in spite of normal V-I (V-I=0.50). 
 Fourier metallicity [Fe/H]=-1.79. 
Photometric calibration error (See Sect. \ref{s:cmd}). 
 \item V125: Shifted in the Av-logP plane towards long periods. V magnitude
 $\sim$ 0.1 brighter than the bulk M54 RRL population. Probably evolved
 variable. 
 \item V128: bright. 
Normal location in the Av-logP plane. Located within the central 1 arcmin, 
possibly a blend with a field ``blue plume'' star. 
 \item V130: V magnitude $\sim$ 0.2 mag fainter than the bulk M54 RRL population.
 Normal location in the Av-logP plane. 
 \item V132 bright, and redder than the assumed red edge of the instability strip. 
Normal location in the Av-logP plane. 
Located within the central 1 arcmin, possibly a blend with a ``red clump'' star. 
 \item V133: unusually blue B-V colour (B-V = 0.18). 
Photometric calibration error (See Sect. \ref{s:cmd}). 
 \item V137: Shifted in the Av-logP plane towards long periods. V magnitude
 $\sim$ 0.1 brighter than the bulk M54 RRL population. Probably evolved
 variable. 
 \item V139: the period-amplitude values locate the star among the RR1 variables, 
however the light curve shows the typical shape of a fundamental pulsator. 
The metallicity obtained from the Fourier parameters is [Fe/H]=-0.04. 
Therefore this star likely belongs to the metal-rich ($\sim$ solar metallicity) 
tail of the Sgr field stellar population. 
 \item V140: Shifted in the Av-logP plane towards long periods. Normal magnitude.
 Possibly belonging to Sgr. 
 \item V141: Metal-rich ([Fe/H]=-1.13). Normal V magnitude and position in the
 Av-logP diagram. Possible Sgr member.
 \item V142: bright. Located within central 1 arcmin, 
possibly a blend. 
 \item V143: unusually red colour (B-V = 0.99). 
Photometric calibration error (See Sect. \ref{s:cmd}). 
\item V145: the classification of this star is highly uncertain. The data from 
the first two night would be compatible with a RR1, those from the third night
 are not. The period and the complete light curve may suggest a classification 
 as a RS CVn variable, with a possible alternative period of 1.053375 days.  
Its position in the CMD (straight on the cluster HB) may indicate a similarity 
with the case of the superposition of two RRL discussed by 
Corwin \& Carney (1998).
\item V147: period and the B light curve are quite typical of RR0 stars, with an 
amplitude of $\simeq 0.6$ mag. The V light curve does not seem compatible with 
this interpretation, probably due to an error in the photometric zero point 
(from DoPHOT photometry of the V reference frame). For this reason we flag this
 variable as uncertain.
\item V211: uncertain classification.
\end{itemize}

\label{lastpage}

\end{document}